\documentclass[jcp,onecolumn,preprint,floatfix]{revtex4-1}

\usepackage{graphics} 
\usepackage{graphicx} 
\usepackage{color} 
\usepackage{bm} 
\usepackage{bbm}
\usepackage[dvipsnames]{xcolor,colortbl}
\usepackage{amsmath}
\usepackage{amssymb}
\usepackage{amsfonts}
\usepackage[normalem]{ulem}  

\definecolor{DarkYellow}{RGB}{80, 80, 0}

\let\vec\boldsymbol

\usepackage{amsmath,amssymb,graphicx}
\usepackage{amsbsy}
\usepackage{latexsym}
\usepackage{xcolor}
\usepackage{graphicx}
\usepackage{psfrag}
\usepackage[normalem]{ulem}
\usepackage{bm}
\usepackage{hyperref}
\hypersetup{colorlinks=true, linkcolor=red, citecolor=blue, urlcolor=blue}
\usepackage{float}
\usepackage{multirow}
\begin{document} 

\title{Two-stage assembly of patchy ellipses: From bent-core particles to liquid crystal analogs}
\author{Anuj Kumar Singh$^{1}$}
\author{Arunkumar Bupathy$^{1,2}$}
\author{Jenis Thongam$^{3}$} 
\author{Emanuela Bianchi$^{3,4}$} 
\author{Gerhard Kahl$^{3}$}
\author{Varsha Banerjee$^{1}$}

\affiliation{$^1$Department of Physics, Indian Institute of Technology Delhi, New Delhi 110016, India.}
\affiliation{$^2$School of Chemistry, University of Birmingham, Birmingham B15 2TT, UK.}
\affiliation{$^3$Institut f\"{u}r Theoretische Physik, TU Wien, Wiedner Hauptstra{\ss}e 8-10, A-1040 Wien, Austria}
\affiliation{$^4$CNR-ISC, Uos Sapienza, Piazzale A. Moro 2, 00185 Roma, Italy.}


\begin{abstract}
We investigate the two-dimensional behavior of colloidal patchy ellipsoids specifically designed to follow a two-step assembly process from the monomer state to mesoscopic liquid-crystal phases, via the formation of so-called bent-core units at the intermediate stage. Our model comprises a binary mixture of ellipses interacting via the Gay-Berne potential and decorated by surface patches, with the binary components being mirror-image variants of each other - referred to as left-handed and right-handed ellipses according to the position of their patches. The surface patches are designed so as in the first stage of the assembly the monomers form bent-cores units, i.e. V-shaped dimers with a specific bent angle. The Gay-Berne interactions, which act between the ellipses, drive the dimers to subsequently form the characteristic phase observed in bent-core liquid crystals. We numerically investigate -- by means of both Molecular Dynamics and Monte Carlo simulations -- the described two-step process: we first optimize a target bent-core unit and we then fully characterize its state diagram in temperature and density, defining the regions where the different liquid crystalline phases dominate.
\end{abstract}
\date{\today}
\maketitle

\section{Introduction}
\label{sec:introduction}

During the past years bent-core liquid crystals (BCLCs) have attracted a rapidly increasing interest both in experimental and theoretical investigations \cite{Lubensky2002, Ros2005, Takezoe2006, Jakli2018}. The constituent units of these systems are bent-core (BC) colloidal entities (typically ranging in size from nm to $\mu$m) that have a (sharp or smooth) bent shape. This characteristic shape can be realized in different ways, for instance by merging two ellipsoids (or sphero-cylinders) via their tips at a (possibly flexible) angle~\cite{Johnston2002, Bates2005, Bates2006} or by creating bead-resolved, bent strings of spherical particles (often referred to as ``banana-shaped'' colloids)~\cite{Dewar2004, Dewar2005, Chiappini2019}. The reason for the considerable interest in such BC particles is the large variety of (ordered) phases that these entities can form: fascinating chiral nematic, smectic, splay, and twist-bend phases (despite the achiral nature of the BC constituents) are just a few examples. 

Here we propose a model system where the BC units result from the spontaneous assembly of suitably designed particles characterized by shape and bond anisotropy, namely we design anisotropic, patchy colloids such that they form V-shaped dimers. To introduce shape anisotropy, we choose -- in our two-dimensional setup -- elliptic particles, which interact via the well-established Gay-Berne (GB) potential~\cite{Berardi1995}. The relevant parameters that characterize this interaction are the aspect ratio of the ellipses and their interaction range. The other ingredient to our model is the patchiness of the colloids~\cite{Kretzschmar_2010,Bianchi_pccp_2011,ravaine2017synthesis,Bianchi2017Limiting}: patchy colloids have surfaces decorated with well defined regions (both in terms of position and of spatial extent) that support a different interaction behaviour with respect to the bare surface. Thus, depending on their mutual interaction properties these regions can either foster or suppress patch-patch directional bonding. Patchiness has become a widely used standard tool (both in experiments and in theoretical/simulation-based investigations) to trigger in bottom-up processes the self-assembly of decorated particles into highly complex bonding patterns, ranging from liquid-crystalline mesophases, over complex ordered structures and quasi-crystalline lattices to structural glasses~\cite{SmallenburgLeibler_2013, Granick_2011, Pine_2012, Chakrabarti_2018, Pine_2020, Liedl_2024, Sulc_2024, karner2024anisotropic}.

In the present contribution we describe in detail the process that starts from the formation of BCs via the self-assembly of elliptic patchy colloids and ends up in the self-organization of these entities into BCLCs. Our design of the elliptic patchy particles envisages a binary mixture of mirror-symmetric units: while both species have the same elliptic shape, the two types of particles are mirror symmetric with respect to their patch decoration (leading thus to left- and right-handed particles). In fact, the two patches (per particle) are located close to the tip of the colloid. The form of the patch-patch interaction is based on the work of Russo et al. \cite{Russo2009}, while the choice of the position, interaction range and interaction strength of the patches influences the features of the BC units.
In a first step we demonstrate that our particles are indeed able to self-assemble via the related patch-patch bonding mechanism into stable BC entities; we demonstrate that our model is characterized by a high versatility, as the bonding angle can be tailored via the patch positions and/or the patch-patch interaction strength. The impact of all these parameters on the actual shape, stability and yield of these units is studied in detail. In a second step we consider an ensemble of such BCs and study their self-assembly into mesophases. Focusing for the moment on a particular set of model parameters we are able to trace out in a qualitative manner a diagram of states (in the temperature-density plane), which shows the emergence of isotropic and smectic phases.  All investigations have been carried out via extensive computer-simulations: we have applied both molecular dynamics (MD) and -- in a complementary manner -- Monte Carlo (MC) simulations, using different types of thermodynamic ensembles. As the proposed BC model can be easily extended to support BCs with different sets of features, it reveals itself as a powerful tool to gain a deeper insight into this fascinating class of functional materials with potential applications in optics, photonics, and sensing. The rest of our paper is organized as follows. Sec.~\ref{sec:model} provides the theoretical framework which includes the model and the methodology to characterize the bent-cores. The computational details of the MD and MC procedures are provided in Sec.~\ref{sec:computational_details}. The results and discussion follow in Sec.~\ref{sec:results}. Finally, Sec.~\ref{sec:conclusions} provides the summary and conclusion.

\section{Theoretical framework}
\label{sec:model}
\subsection{Model}

We want to design a simple model in which colloidal monomers form a bent-core (BC) entity first, which then self-assemble into mesoscopic liquid crystalline phases. Such a targeted design process would allow for exploring a large number of phases which emerge by changing the shape as well as the surface chemistry of the colloidal monomers. 

To simulate such a system we use a simple Gay-Berne (GB) potential \cite{Berardi1995} for the elliptic colloidal monomers which are decorated on their surface by suitably placed patches (index `P'). The form of the patch potential is adopted from the work of Russo \textit{et al}.~\cite{Russo2009}. Thus the interaction potential between a pair of particles $i$ and $j$  is given by (where we have suppressed for the moment the arguments)
\begin{equation}
V_{ij} = V_{\rm GB} + V_{\rm P},
\label{eq:model}
\end{equation}
where $V_{\rm GB}$ is the Gay-Berne potential acting between the centers of mass of the two particles, and $V_{\rm P}$ is the interaction potential due to all pairs of patches located on the surfaces of particles $i$ and $j$.

\begin{figure}[H]
\centering
\includegraphics[width=1\textwidth]{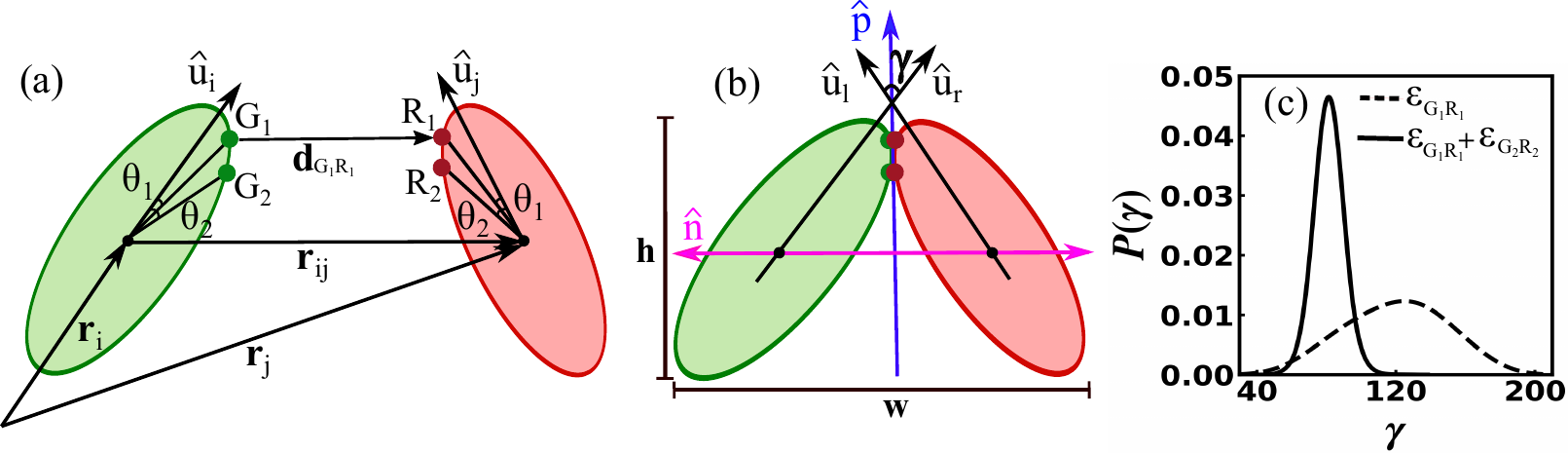}
\caption{(a) Schematic representation of the patchy colloidal model used in this contribution with right-handed (green) and left-handed (red) elliptic particles with positions $\vec{r}_i$, $\vec{r}_j$ and orientations $\hat{u}_i$, $\hat{u}_j$. The right-handed particle has patches $G_1$ and $G_2$ at angles $-\theta_1$ and $-\theta_2$ and the left-handed particle have patches $R_1$ and $R_2$ at angles $-\theta_1$ and $-\theta_2$ (as indicated). $r_{ij} = |\vec{r}_j - \vec{r}_i|$ represents the separation between the center of masses of the two particles, and $d_{G_1R_1}$ the separation between patches $G_1,R_1$. (b) Schematic depiction of a BC particle having nematic $\hat{n}$ and polar $\hat{p}$ directions and bend angle $\gamma$. (c) Distribution of bend angles $P(\gamma)$ vs. $\gamma$ obtained using single patch particles (dashed line) and two patch particles (bold line) for $\epsilon_p=15$, $\alpha=0.1$, $\theta_1=5$, $\theta_2=10$, at $\rho=0.4$ and $T=1.2$.}
\label{fig:model}
\end{figure}

The GB potential $V_{\rm GB}$ characterizes the shape-anisotropy of the elliptical particles and is defined as follows:
\begin{multline}
V_{GB}(\hat{u}_i, \hat{u}_j, \vec{r}_{ij}) = 4\epsilon(\hat{u}_i, \hat{u}_j,\hat{r}_{ij})\left[\left(\frac{\sigma_0}{r_{ij} - \sigma(\hat{u}_i, \hat{u}_j, \hat{r}_{ij})+\sigma_0}\right)^{12} \right. \\
\left. - \left(\frac{\sigma_0}{r_{ij} - \sigma(\hat{u}_i, \hat{u}_j, \hat{r}_{ij})+\sigma_0}\right)^{6}\right],
\label{eq:GayBerne}
\end{multline}
where $\sigma_0$ is the unit distance of the interactions.  $\vec{r}_{ij}$ is the vector connecting the centers of mass of the particles and $\hat{u}_i$ and $\hat{u}_j$ are the unit vectors representing the orientations of the particles as shown in Fig.~\ref{fig:model}(a). $\epsilon = \epsilon_0 \epsilon_1^{\mu} \epsilon_2^{\nu}$ is the orientation dependent interaction strength, where $\epsilon_0$ represents the energy unit of the interactions; the parameters $\mu$ and $\nu$ modify the well depth of the potential; $\sigma$ is the anisotropic contact distance. Further $\epsilon_1$, $\epsilon_2$ and $\sigma$ are defined as:
\begin{align}
    \epsilon_1(\hat{u}_i, \hat{u}_j, \hat{r}_{ij}) &= 1 - \frac{\chi'}{2}\left\{\frac{(\hat{r}_{ij}.\hat{u}_i + \hat{r}_{ij}.\hat{u}_j)^2}{1+\chi'\hat{u}_i . \hat{u}_j}+\frac{(\hat{r}_{ij}.\hat{u}_i - \hat{r}_{ij}.\hat{u}_j)^2}{1 - \chi'\hat{u}_i . \hat{u}_j}\right\}, \\
    \epsilon_2(\hat{u}_i, \hat{u}_j) &= \left[1-\chi^2(\hat{u}_i . \hat{u}_j)^2\right]^{-\frac{1}{2}},\\
    \sigma(\hat{u}_i, \hat{u}_j, \hat{r}_{ij}) &= \sigma_0\left[1-\frac{\chi}{2}\left\{\frac{(\hat{r}_{ij}.\hat{u}_i + \hat{r}_{ij}.\hat{u}_j)^2}{1+\chi\hat{u}_i . \hat{u}_j}+\frac{(\hat{r}_{ij}.\hat{u}_i - \hat{r}_{ij}.\hat{u}_j)^2}{1 - \chi\hat{u}_i . \hat{u}_j}\right\}\right]^{-\frac{1}{2}}.
\label{eq:GBStrengthRange}
\end{align}
The energy anisotropy parameter $\chi'$ and the shape anisotropy parameter $\chi$ are defined as:
\begin{align}
    \chi' &= (\kappa'^{1/\mu}-1) / (\kappa'^{1/\mu} + 1), \\
    \chi &= (\kappa^2-1) / (\kappa^2 + 1), 
\end{align}
where $\kappa' = \epsilon_s / \epsilon_e$ is the energy anisotropy with $\epsilon_s$ and $\epsilon_e$ being the well depths of the side-to-side and end-to-end configurations. $\kappa = \sigma_{\parallel} / \sigma_{\perp}$ is the aspect ratio of the elliptic particles with $\sigma_{\parallel}$ and $\sigma_{\perp}$ being the major and minor axes of the particles. The GB potential therefore contains four key parameters to describe interacting elliptic particles: $\kappa$, $\kappa'$, $\mu$, and $\nu$. There exist a number of GB homologues in the context of nematic LCs depending on the values chosen for the four parameters ~\cite{Gay1981, Berardi1993, Bates1999, Zannoni2001}. We will discuss the appropriate choice for obtaining the BCs and their LC-like self-assembly shortly.

The patch-patch interactions $V_{\rm P}$ are described by a sum of inverted Gaussian potentials~\cite{Russo2009}:
\begin{equation}
V_{\rm P}(d) = - \sum_{k \in P_i} \sum_{l \in P_j} \epsilon_{kl} \exp\left[-\frac{1}{2}\left(\frac{d_{kl}}{\alpha}\right)^2\right],
\label{eq:invGauss}
\end{equation}
where $\epsilon_{kl}$ is the strength of the interaction between patches $k$ and $l$, $P_i$ and $P_j$ are the set of patches on the particles $i$ and $j$, respectively, and $d_{kl}$ is the distance between the centers of patches $k$ and $l$, which decorate the surface of the elliptical particles, as shown in the schematic of Fig.~\ref{fig:model}(a). The width of the inverted Gaussian, $\alpha$, defines the range of the patch-patch interactions. In addition to being able to compute gradients required for molecular dynamics (MD) simulations, this choice of the patch potential also allows for oblique patch placements, \textit{i.e.}, at an angle from the tip of the ellipses. Such placements promote the assembly of stable BCs. 

Let us consider the schematic in Fig.~\ref{fig:model}(a) that shows ellipsoidal particles with two patches placed obliquely; the reason for a two-patch model will be clear soon. In our model there are two kinds of patch placements on the ellipsoids: one to the ``left'' (red particles) and the other to the ``right'' (green particles). The green particles have patches $G_1$ and $G_2$ at angles $-\theta_1$ and $-\theta_2$ from the major axis. Similarly, the red particles have patches $R_1$ and $R_2$ at angles $\theta_1$ and $\theta_2$ from the major axis. The interaction strengths $\epsilon_{G_1 R_1}$ and  $\epsilon_{G_2 R_2}$ are set equal to $\epsilon_p$, while all other interactions are set to 0: these include notably $\epsilon_{G_1G_1}$, $\epsilon_{G_1G_2}$, $\epsilon_{R_1R_1}$, $\epsilon_{R_1R_2}$, $\epsilon_{G_1 R_2}$ and $\epsilon_{G_2 R_1}$. We consider a pair of left-handed and right-handed particles to be a BC unit if the patch interaction strength $V_p$ is 
sufficiently large to prevent dissociation of this arrangement by thermal fluctuations. In our simulations, we find that a choice of $V_p < -3k_{\rm B}T$  ($T$ being the temperature and $k_{\rm B}$ being the Boltzmann constant) ensures that the BCs once formed remain bonded.  

Fig.~\ref{fig:model}(b) shows a schematic of a BC colloid corresponding to the above model description. This entity is characterized by the director ${\hat{n}}$, the polarization $\hat{p}$, and the bend angle $\gamma$ (as indicated). In our investigations we have found that single patch particles are characterized by a broad distribution of the bend angles, as shown in Fig.~\ref{fig:model}(c). Reducing the width of the  angle distribution would require lowering the value of $\alpha$, which, as we shall see later, negatively affects the kinetics of BC self-assembly. Further, multiple single-patch ellipses can adhere in clusters (or asters). 
Having two patches on the other hand results in smaller magnitudes of the bend angle and lesser flexibility.  
Thus the two-patch ellipsoids represent an optimal choice to obtain BCs at a high yield. 
The model parameters $\theta_i$, $\epsilon_p$, and $\alpha$ have a significant impact on the BC yield and phases. We will study their influence in the rest of the paper.

\subsection{Methods}
\label{subsec:methods}

\subsubsection{\textbf{Bent-core characteristics}}

Let us refer to the schematic in Fig.~\ref{fig:model} that depicts a pair of left-handed and right-handed patchy ellipsoids. Let their center of mass, after the formation of the BC, be represented by $\vec{r}_{\rm cm}^{\rm l}$ and $\vec{r}_{\rm cm}^{\rm r}$, respectively. The center of mass of the BC unit is given by the vector average:

\begin{equation}
\label{CM}
   \vec{R}_{\rm cm} = \frac{\vec{r}_{\rm cm}^{\rm l} + \vec{r}_{\rm cm}^{\rm r}}{2}.
\end{equation}
Further, the local director $\hat{n}$ and the polarization $\hat{p}$ can be expressed in terms of the orientations $\vec{u}_l$ and $\vec{u}_r$ as follows:

\begin{equation}
\label{np}
    \hat{n}= \frac{\vec{u}_{\rm l}-\vec{u}_{\rm r}}{|\vec{u}_{\rm l}-\vec{u}_{\rm r}|}, \quad  \hat{p} =\frac{\vec{u}_{\rm l}+\vec{u}_{\rm r}}{|\vec{u}_{\rm l}+\vec{u}_{\rm r}|},
\end{equation}
Similarly, the bend angle $\gamma$ can also be defined in terms of the orientation vectors as follows:

\begin{equation}
\label{gamma}
\gamma=\cos^{-1}\left( \frac{\vec{u}_{\rm l} \cdot \vec{u}_{\rm r}}{|\vec{u}_{\rm l}||\vec{u}_{\rm r}|}\right).
\end{equation}
As discussed in the context of Fig.~\ref{fig:model}(b), the BC unit is characterized by $\hat{n}$, $\hat{p}$, and $\gamma$. We will use Eqs.~(\ref{np}) and (\ref{gamma}) for the evaluation of these quantities in all our simulations. 

\subsubsection{\textbf{Q-Tensor}}
\label{Qtensor}

To characterize the global orientational order of the mesoscopic phases formed by the BC units, we compute the $Q$-tensor as an appropriate order parameter for quantifying both the nematic and the polar order~\cite{De1993physics, Wei1992, Allen2017}. Its components are defined as follows:

\begin{equation}
Q_{\alpha\beta} = \frac{1}{N_{\rm b}}\sum_{i=1}^{N_{\rm b}}\frac{1}{2}\left(3 n_{i,\alpha}n_{i,\beta} - \delta_{\alpha \beta}\right),
\end{equation}
where $N_{\rm b}$ is the number of BCs in the ensemble, $\alpha $, $\beta \in \{x,y\}$ represent the Cartesian components of $\hat{n}_i$ corresponding to the $i^{th}$ BC. The largest eigenvalue of $Q$ gives the nematic order parameter $\mathcal{S}$ and the corresponding eigenvector represents the global director ${\bf n}$. The nematic order parameter can also be interpreted as $\mathcal{S}=\overline{P_2(\hat{n}_i\cdot\mathbf{n})}$, where $P_2(x) = (3x^2-1)/2$ is the Legendre polynomial of degree two and the over-bar indicates an average over all the molecules. The isotropic phase corresponds to ${\cal S} = 0$ and a fully aligned nematic phase has ${\cal S} = 1$. 
The polar order can be computed as $\mathcal{P} = \left|\overline{\hat{p}_i\cdot\mathbf{p}}\right|$, where $\mathbf{p}$ is normal to $\mathbf{n}$. A polarization $\mathcal{P} = 1$ implies that the polar vectors $\hat{p}_i$ of the BCs are all pointing in the same direction as $\mathbf{p}$. In this case, aggregates of BCs exhibit nematic as well as polar order. $\mathcal{P} = 0$ indicates random or staggered (\textit{e.g.}, anti-ferroelectric) orientation of $\hat{p}_i$ of the BCs. Note that in this case, the aggregate could still exhibit nematic order.

\subsubsection{\textbf{Radial distribution function}}
\label{PCF}

The radial distribution function $g(r)$ (RDF) which measures correlations between pairs of particles can provide information about the presence of density modulations in the system and is a useful tool for identifying the structure of the aggregates or domains. This function is defined as \cite{Frenkel2023,Jeffrey1998}
\begin{equation}
g(r) = \frac{1}{\rho_{\rm b} N_{\rm b}}\left\langle \sum_{\stackrel{i,j}{i\neq j}}^{N_{\rm b}} \delta(r-r_{ij}) \right\rangle,
\end{equation}
where $r_{ij} = |\vec{r}_{ij}|$ is the separation distance between the centers of mass of BCs $i$ and $j$, $N_{\rm b}$ is the number of BCs, and  $\rho_{\rm b} = N_{\rm b}/L^2$ is their number density. In practice, the $\delta$-function is replaced by a counting function which is normalized appropriately by the area of the finite but small region in which the counting is done. The division by $\rho_{\rm b}$ ensures that $g(r) = 1$ when the particles are completely uncorrelated, as this is, for instance realized in an ideal gas. Sharp peaks indicate the presence of structure and oscillatory behavior marks the presence of translational order in the system. We can also define a characteristic lengthscale (or correlation length) over which the particle positions are correlated. Typically, it is the distance $r$ over which the peaks in $g(r)$ decay to -- say -- 0.2 times the maximum value of the RDF \cite{Puri2004}.

In order to identify layered phases such as the smectic one, we resolve the RDF along axes in the local frame of reference of the BC units and average the results. The RDF resolved parallel to the local director $\hat{n}_i$ of the BCs, $g_{\parallel}(r_{\parallel})$, is defined as follows \cite{Weis1993,Caneda2014}:

\begin{equation}
g_{\parallel}(r_{\parallel})=\frac{1}{ \rho_b N_b}\left\langle\sum_{\stackrel{i,j}{i\neq j}}^{N_b}\frac{\delta(r_{\parallel} - r_{ij, \parallel})\Theta(h/2 - r_{ij,  \perp})}{h}\right\rangle,
\end{equation}
where $r_{ij, \parallel}= |\vec{r_{ij}} \cdot \hat{n}_i|$ and $r_{ij,\perp}=|\vec{r_{ij}}\times\hat{n}_i|$ are the parallel and perpendicular components of the separation $r_{ij}$ with respect to $\hat{n}_i$; further, $h$ is the average height of the BCs (\textit{i.e.}, average extension of the BCs in the direction perpendicular to $\hat{n}_i$, see Fig.~\ref{fig:model}b) and $\Theta (x)$ is the Heaviside step function, which ensures that only BCs within the same layer as the reference BC (with index $i$) are included in the computation. Similarly, $g_{\perp}(r_{\perp})$ which resolves RDF in the direction perpendicular to the local director is defined as~\cite{Weis1993, Caneda2014}

\begin{equation}
g_{\perp}(r_{\perp})=\frac{1}{\rho_{\rm b} N_{\rm b}}\left\langle\sum_{\stackrel{i,j}{i\neq j}}^{N_{\rm b}}\frac{\delta(r_{\perp} - r_{ij, \perp})\Theta(w/2 - r_{ij,  \parallel})}{w}\right\rangle,
\end{equation}
where $w$ is the average width of the BCs (see Fig.~\ref{fig:model}b). 

\section{Computational Details}
\label{sec:computational_details}

Our starting point to obtain BC colloids and their phases is a collection of an equal number of right-handed and left-handed two-patch ellipsoids, and specified patch-patch interactions. We study the self-assembly using extensive molecular dynamics (MD) and Monte Carlo (MC) simulations. As discussed in detail in Sec.~\ref{sec:model}, the ellipsoids interact via the GB potential which is characterized by four essential parameters: $\kappa$, $\kappa^\prime$, $\mu$ and $\nu$. There are a large variety of GB homologues which differ from each other in terms of the values chosen for the four parameters. The primary focus in selecting their values has been (i) to obtain the nematic and smectic LC phases and (ii) to obtain convergence with experimental data \cite{Gay1981, Berardi1993, Bates1999, Zannoni2001}. The choice of $\kappa=3.0$ is elementary as for real LC systems, the length-to-breadth ratio of the constituent molecules is typically equal to or greater than 3:1. 
A phase diagram of the GB model in the $\kappa^\prime-T$ indicates that the ordered phase as $\kappa^\prime\rightarrow 1$ is primarily nematic \cite{Birdi2022}. A value of $\kappa^{\prime} > 1$ introduces energy anisotropy, favoring side-to-side alignment of ellipses. These values promote nematic as well as smectic order characteristics of LCs. The parameters $\mu$ and $\nu$ modify the well depths of the potential, and hence their impact on the nematicity and smecticity is very subtle \cite{Gay1981, Berardi1993, Bates1999, Zannoni2001}. A popular choice of parameters is $\kappa=3$, $\kappa^\prime=5$, $\mu=1$ and $\nu=3$ due to Berardi {\it et al.} \cite{Berardi1993}. They ensure the formation of LC  phases and agreement with experimental data in many real systems \cite{Berardi1993}. In the present work, our focus is to first create BC units, and then study their assembly in ordered phases. Consequently, we choose $\kappa^\prime=1$ to promote patch-patch aggregation for the formation of BCs in the first stage, which then assemble to form ordered phases due to the GB interactions in the second stage. 

 
\subsection{Details of MD simulations}
MD simulations are performed in NVT and NPT ensembles using the LAMMPS software~\cite{LAMMPS, Plimpton1995}, employing the Nos\'e-Hoover thermostat and barostat for temperature and pressure coupling, as needed \cite{Frenkel2023, Hoover1985, Hoover1999}. Periodic boundaries with minimum image convention have been used for the simulations. All calculations were done in reduced LJ units: $T^*= k_BT/\epsilon_0$, $ \Delta t^*= \Delta t/ \sqrt{m \sigma_0^2/ \epsilon_0}$, where $k_B$ is the Boltzmann constant. The star is dropped in subsequent discussions. The equations of motion are integrated using the velocity-Verlet algorithm with simulation time step $\Delta t = 0.001$ \cite{Andersen1983}. 

We initialize the system with $N=4000$ patchy ellipsoids in a square box of length $L$ depending on the desired density $\rho=(\pi \sigma_{\parallel} \sigma_{\perp})N/L^2$, where $ \sigma_{\parallel}$  and $ \sigma_{\perp}$ are the major and minor axes of the ellipsoids. For all studied densities, we initialize the particles on a regular lattice and allow the system to equilibrate at $T=3$ for $10^5$ MD steps. The resulting homogeneous state is then quenched to the desired temperature. Simulations were performed up to $10^7$ steps and relevant quantities measured at intervals of $5\times 10^3$ steps. All presented data have been averaged over 10 independent runs.

\subsection{Details of MC simulations}

In a complementary fashion we have used Monte Carlo (MC) \cite{Frenkel2023,Allen2017} simulations to investigate the structural properties of our BC entities. To be more specific we have used so-called ``virtual-move'' MC (VMMC) simulations \cite{VMMC_1, VMMC_2, VMMC_3}, an algorithm that proposes simultaneous translational and rotational moves of clusters of particles according to gradients of interaction energies of the particles. In short, a seed particle is subjected to a ``virtual'' trial move; based on the emerging bond energies before and after the trial move, neighbouring particles are added to a virtual cluster. This process is iterated for each particle in the cluster and the positions of all particles involved are updated simultaneously (for more details we refer to Refs. \cite{VMMC_1, VMMC_2}). 

The VMMC simulations were performed in the NVT ensemble, applying standard periodic boundary conditions. Due to the fact that the degree of parallelization of the VMMC algorithm is strongly limited, we had to restrict ourselves to considerably smaller ensembles (as compared to the MD simulations): a typical ensemble size of $N = 900$ particles was used. Starting in a square box (whose size is imposed by the desired density $\rho$) half of the particles are left- and right-handed, respectively. At a given temperature the system is equilibrated during 4 $\times$ 10$^7$ VMMC steps, while the production part of the simulations extended over 2 $\times$ 10$^7$ VMMC steps. For each state point we considered ten independent MC runs, each of them starting from different initial conditions. For the highest densities we started from ordered arrays of parallel unbounded ellipses, while for the lowest densities we started from random configurations.  All quantities were averaged over 4 $\times$ 10$^3$ independent configurations per single run, adding to  4 $\times$ 10$^4$ per state point.

\section{Results and Discussion}
\label{sec:results}

\subsection{First step assembly stage}
Before studying the mesoscopic phases of the system, it is useful to understand the effect of the model parameters on the formation of BCs. The presented patchy colloidal analog of BCLCs has three model parameters and two system parameters to control the properties of the BCs: patch interaction strength $\epsilon_p$ (defined in units of $\epsilon_0$), patch interaction range $\alpha$, patch positions $\theta_i$, density $\rho$, and temperature $T$. We perform simulations in the $NVT$ ensemble with density $\rho=0.4$ and temperature $T=1.2$, corresponding to an equilibrium isotropic fluid state of the underlying Gay-Berne particles. Unless stated otherwise, the patch positions are fixed at $\theta_1 = 5$ and $\theta_2 = 10$. 

\begin{figure}[H]
\centering
\includegraphics[width=1\linewidth]{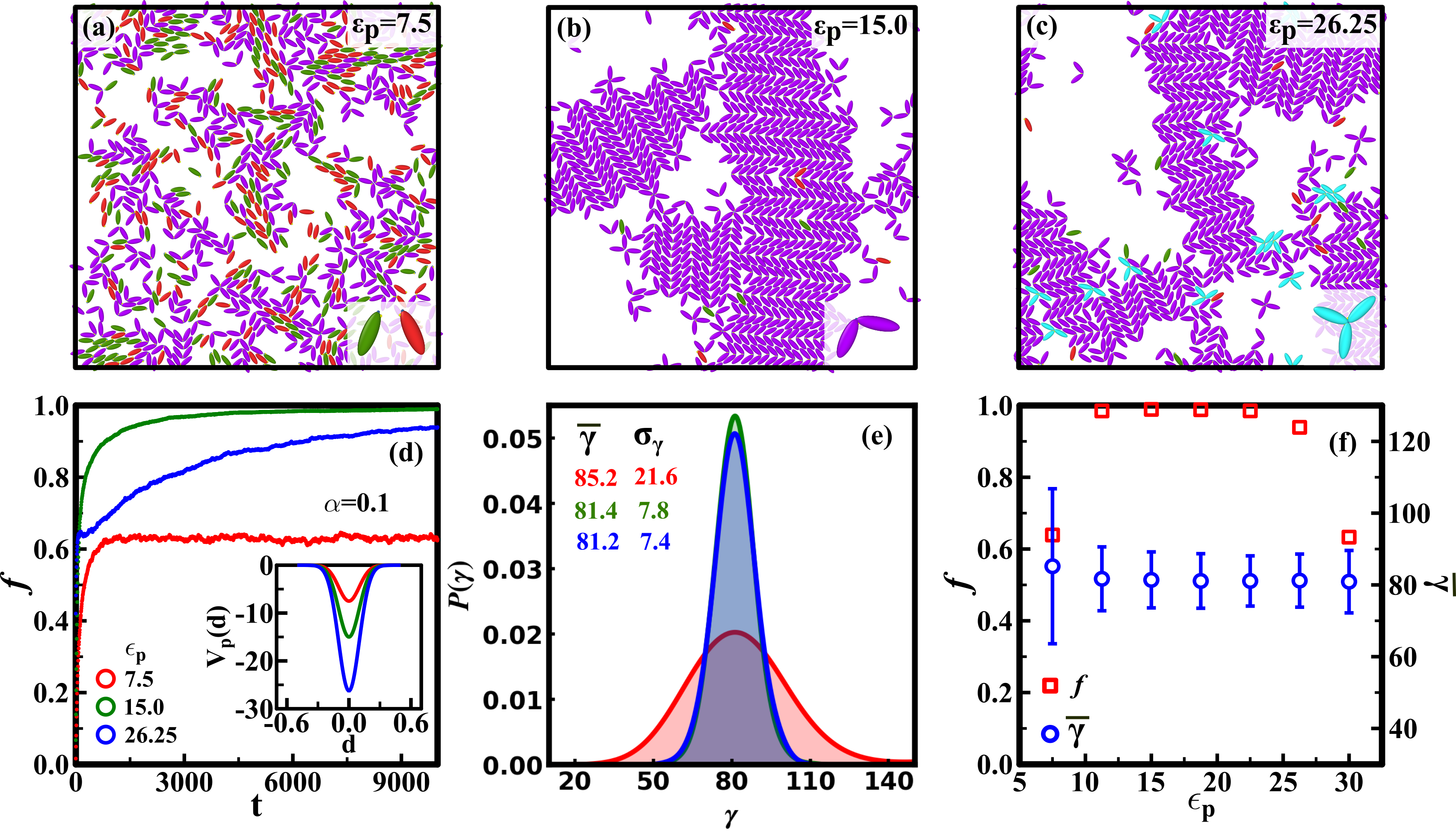}
\caption{Prototypical morphologies at $t=10^4$, $\alpha=0.1$, $T=1.2$ and $\rho=0.4$ for (a) $\epsilon_p=7.5$, (b) $\epsilon_p=15$, and (c) $\epsilon_p=26.25$. In the snapshots, BCs are shown in purple, and asters in cyan. (d) Corresponding plots of BC yield $f$ vs. $t$. The inset depicts the inverted Gaussian potential $V_p$ for $\epsilon_p$ = 7.5, 15, and 26.25 at fixed $\alpha = 0.1$. (e) Distribution of bend angles $P(\gamma)$ vs $\gamma$ with the same color coding as in (d). The average bend angle $\Bar{\gamma}$ and standard deviation $\sigma_{\gamma}$ are mentioned within the plot. (f) Plots of $f$ and $\Bar{\gamma}$ as a function of patch interaction strength $\epsilon_p$ at the end of the simulations. The error bars show the standard deviation of the bend angle.}
\label{fig:epsilon_NVT}
\end{figure}

We start by demonstrating the effect of patch-patch interaction strength $\epsilon_p$ on the formation of BCs, keeping the patch range constant at $\alpha = 0.1$. The important metrics to characterize the BCs are: (i) the fraction of particles $f$ that have self-assembled into BCs, (ii) the average bend angle $\bar{\gamma}$ of the BCs and (iii) the flexibility of the BCs. Figs.~\ref{fig:epsilon_NVT}(a)-(c) show prototypical equilibrated snapshots at time $t=10^4$ for $\epsilon_p$ = 7.5, 15 and 26.25. The red and green patchy ellipsoids which yield a BC unit are coloured purple. The asters, formed when more than two ellipsoidal particles adhere, are coloured cyan. (See insets in Figs.~\ref{fig:epsilon_NVT}(a)-(c).)  The BC yield $f$ for different values of $\epsilon_p$ is shown in Fig.~\ref{fig:epsilon_NVT}(d). The inset shows the patch-patch interaction $V_p(d)$ (inverted Gaussian) for the different values of $\epsilon_p$. The primary effect of varying $\epsilon_p$ is to vary the well-depth. Fig.~\ref{fig:epsilon_NVT}(e) shows that the corresponding distribution of bend angles while Fig.~\ref{fig:epsilon_NVT}(f) shows $f$ and $\bar{\gamma}$ as a function of $\epsilon_p$.
For $\epsilon_p = 7.5$, $f$ remains quite low even after long simulation times. This is because $V_p$ in this case is weaker than thermal energy $k_BT$ and is not sufficient to keep the left-handed and right-handed particles bonded, see the morphology in Fig.~\ref{fig:epsilon_NVT}(a). Fig.~\ref{fig:epsilon_NVT}(e) shows that the corresponding distribution of bend angles is broad.
For an intermediate value of $\epsilon_p = 15$, $V_p$ is sufficiently strong with respect to $k_BT$, thus keeping the individual building blocks bonded, as shown in Fig.~\ref{fig:epsilon_NVT}(b). As a result, we achieve a BC yield close to $1$.
For $\epsilon_p=30$, $V_p$ is even stronger, making single bonds also stable against thermal fluctuations. In addition to the BCs, this leads to the creation of asters formed due to the patch at $\theta_1=5$. Single bonds with more than one monomer of the opposite handedness are also stable in this case as seen in the morphology of Fig.~\ref{fig:epsilon_NVT}(c). Consequently, the BC yield is also reduced, as observed in Fig.~\ref{fig:epsilon_NVT}(d). The corresponding bend angle distribution shown in Fig.~\ref{fig:epsilon_NVT}(e), which excludes the asters, is narrower than that for $\epsilon_p = 7.5$, indicating the reduced flexibility of the BCs, as is expected with stronger patch-patch interactions. Interestingly, for the range of $\epsilon_p$ values over which the BC formation is efficient, we do not notice a significant variation in the flexibility of the bend angle as revealed by Fig.~\ref{fig:epsilon_NVT}(f).

\begin{figure}[H]
\centering
\includegraphics[width=1\textwidth]{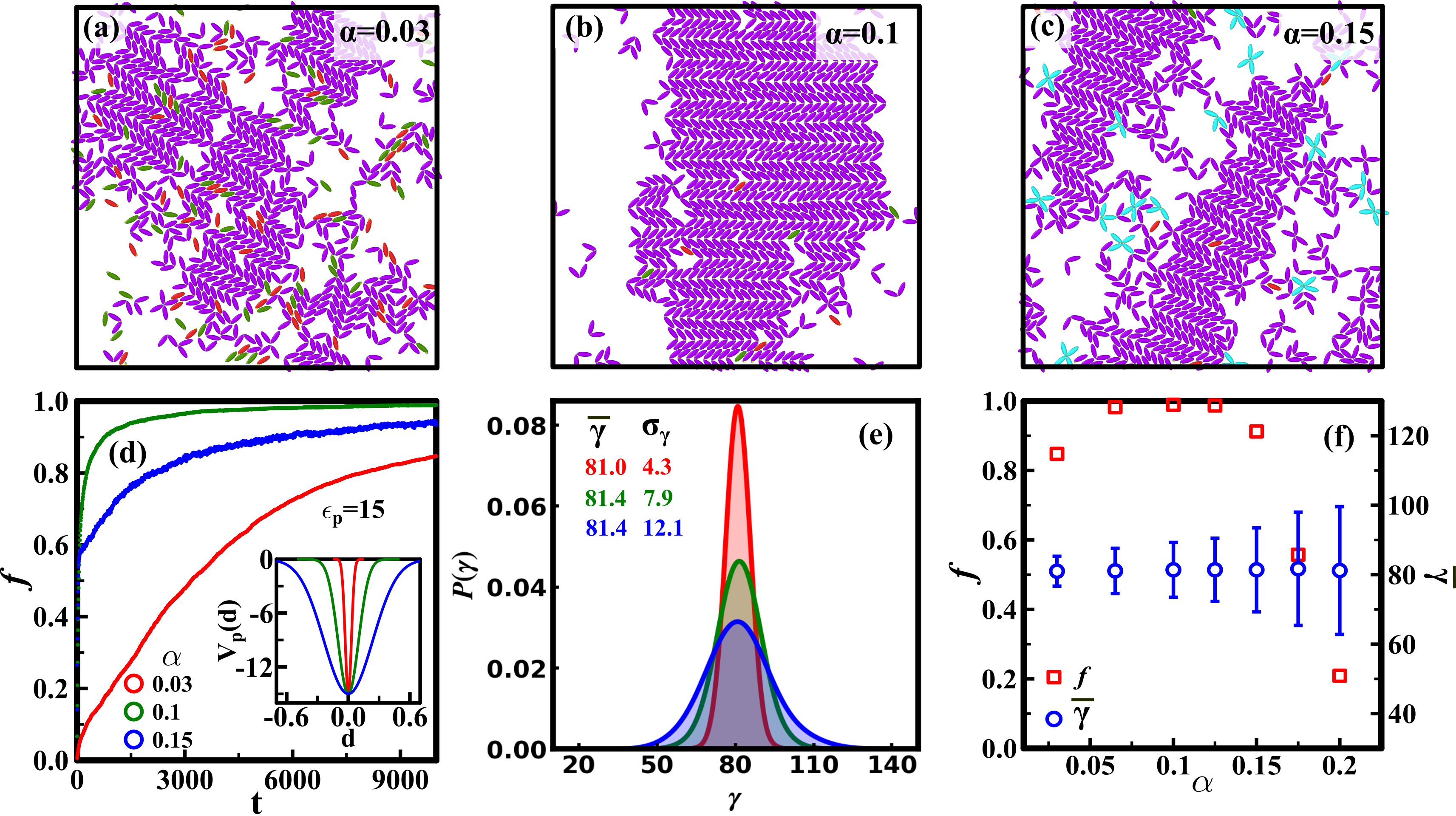}
\caption{Prototypical morphologies at $t=10^4$, $\epsilon_p=15$, $T=1.2$ and $\rho=0.4$ for (a) $\alpha=0.03$, (b) $\alpha=0.1$, and (c) $\alpha=0.15$. (d) Corresponding plots of BC yield $f$ vs. $t$. The inset depicts the inverted Gaussian potential $V_p$ for $\alpha$ = 0.03, 0.1, and 0.15 at fixed $\epsilon_p = 15$. (e) Distribution of bend angles $P(\gamma)$ vs $\gamma$ following the same color coding as panel (d), with the corresponding $\Bar{\gamma}$ and $\sigma_{\gamma}$ indicated within the plot. (f) Plots of $f$ and $\Bar{\gamma}$ as a function of $\alpha$ at the end of the simulations.}
\label{fig:alpha_NVT}
\end{figure}

The other crucial parameter for controlling the self-assembly of BCs is the patch interaction range $\alpha$. Graphically, it represents the well width of inverted Gaussian potential. 
Guided by the observations in Fig.~\ref{fig:epsilon_NVT}, we set the patch strength $\epsilon_p=15$ to explore the effect of $\alpha$. Figs.~\ref{fig:alpha_NVT}(a)-(c) show the snapshots for $\alpha$ = 0.03, 0.1 and 0.15.
For the small value of 0.03, the effective region within which two complementary patches can make a stable bond is significantly reduced. This not only requires the bonding monomers to be close to each other but also that they are in near-perfect alignment in order to make a double bond. As a result, the probability of opposite-handed particles forming a BC is quite small. This negatively affects the kinetics of the self-assembly leading to a low rate of BC production, as seen in Fig.~\ref{fig:alpha_NVT}(d). The resultant BCs are also less flexible and prone to dissociate into monomers upon the smallest fluctuations. 
As $\alpha$ increases to 0.1 in Fig.~\ref{fig:alpha_NVT}(b), the monomers have a higher probability of making bonds and the rate of BC self-assembly increases. The flexibility of BCs also increases as the increased patch range allows for a larger window of angles over which the monomers can remain bonded, as shown in Fig~\ref{fig:alpha_NVT}(e) and (f). Both these factors contribute to a larger BC yield. For larger patch ranges $\alpha \geq 0.15$ in Figs.~\ref{fig:alpha_NVT}(f), the rate of self-assembly of BCs is high. However, the yield is significantly reduced due to the formation of asters. They coexist with the BCs as shown in Fig~\ref{fig:alpha_NVT}(c), and the BCs display a large fluctuation in their bend angle for larger values of $\alpha$. 
From Figs.~\ref{fig:epsilon_NVT}(f) and \ref{fig:alpha_NVT}(f), it can be seen that for fixed values of the patch angles, there is a sizable region in the ($\epsilon_p-\alpha$) parameter space which yields an efficient formation of BCs. The distribution of the bend angle $\gamma$ is predominantly dictated by the patch range $\alpha$ and to a lesser extent by the patch strength $\epsilon_p$. 

\begin{figure}[H]
\centering
\includegraphics[width=1\textwidth]{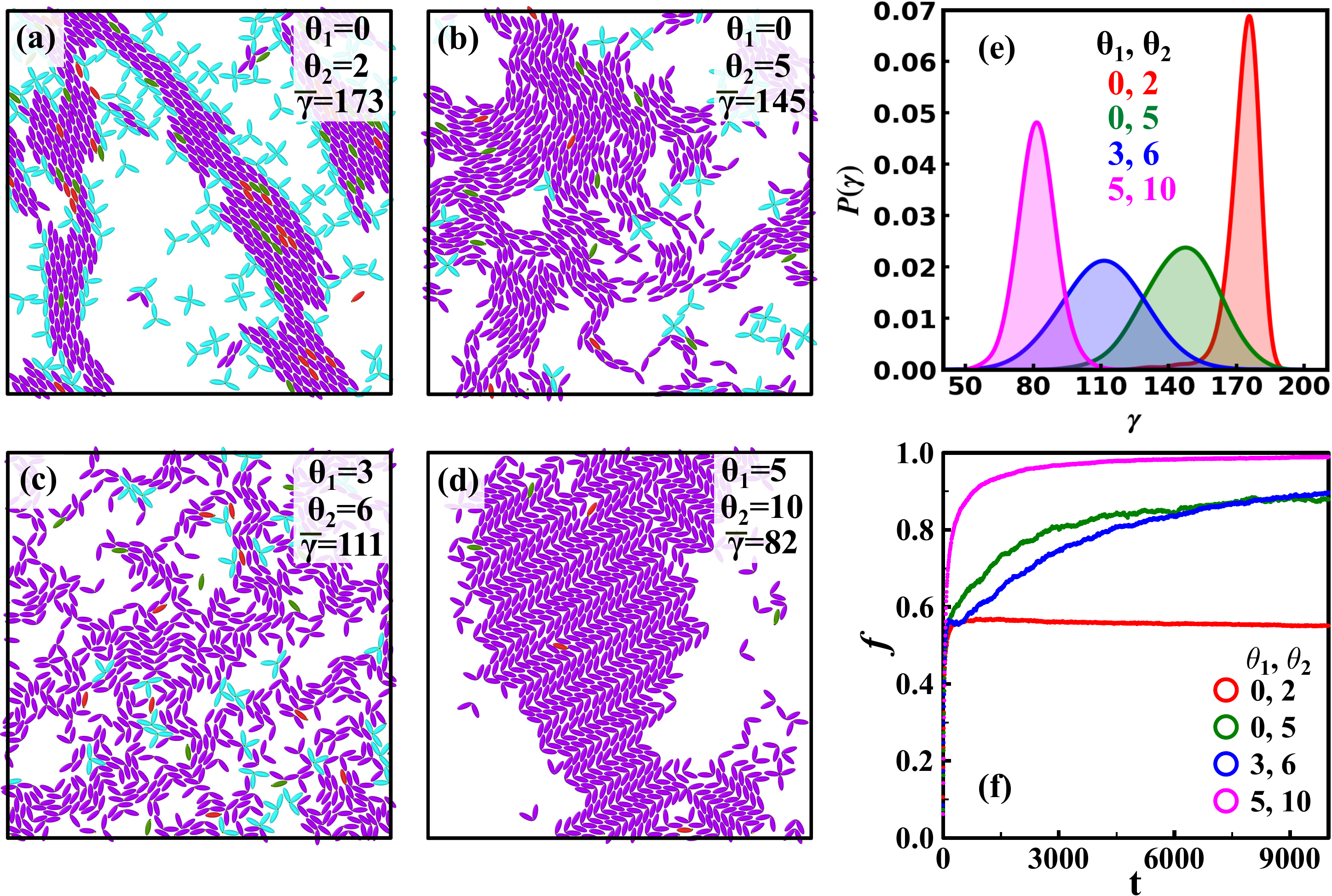}
\caption{Prototypical morphologies at $\epsilon_p=15$, $\alpha=0.1$, $T=1.2$ and $\rho=0.4$ for different patch positions: (a) $\theta_1=0$, $\theta_2=2$, (b) $\theta_1=0$, $\theta_2=5$, (c) $\theta_1=3$, $\theta_2=6$, and (d) $\theta_1=5$, $\theta_2=10$. (e) Corresponding bend angle distributions $P(\gamma)$ vs $\gamma$. (f) Corresponding BC yields $f$ vs. $t$.}
\label{fig:theta}
\end{figure}

It is also educative to understand the role of the patch placements $[\theta_1$, $\theta_2]$ on the BCs. Choosing the $\epsilon_p = 15$ and $\alpha = 0.1$, Fig.~\ref{fig:theta} shows typical morphologies for four different choices of patch positions: (a) [0, 2]; (b) [0, 5]; (c) [3, 6]; (d) [5, 10]. The corresponding bend angle distributions $P(\gamma)$ vs. $\gamma$ and the BC yield $f$ are also shown in Figs.~\ref{fig:theta}(e) and (f). Their choice affects the bend angle. For example [0, 5] yields $\bar{\gamma}\simeq 145$, but shifted placements [5, 10] results in $\bar{\gamma}\simeq 82$. Note that this angle is dependent on the shape anisotropy $\kappa$ as well. For instance, the mean bend angle would be larger for less eccentric ellipses, i.e., for $\kappa < 3$. BC yields are also affected as in cases such as [0, 2], there is a larger possibility of non-BC aggregations such as the asters. Further, we can also expect distinct phases and their textures as the patch positions are varied: [0, 2] nudges the arrangement in a nematic phase while [0, 5] introduces a splay in the texture. Our preferred choice has been [5, 10] as it yields BC formation $\simeq 1$. The interplay of $\alpha$, $\theta_1$ and $\theta_2$ can be expected to yield a range of bend angles and novel phases. This is an exhaustive investigation that deserves attention. However, first it is essential to understand the methodology to create model BCs from patchy ellipsoids and subsequently study their self-assembly into states with long-range liquid crystalline order. We prioritize establishing these protocols in the present work and will address the $\alpha$-$\theta_1$-$\theta_2$ state diagram in an ensuing study.

\subsection{Second step assembly stage}
\begin{figure}[H]
\centering
\includegraphics[width=1\textwidth]{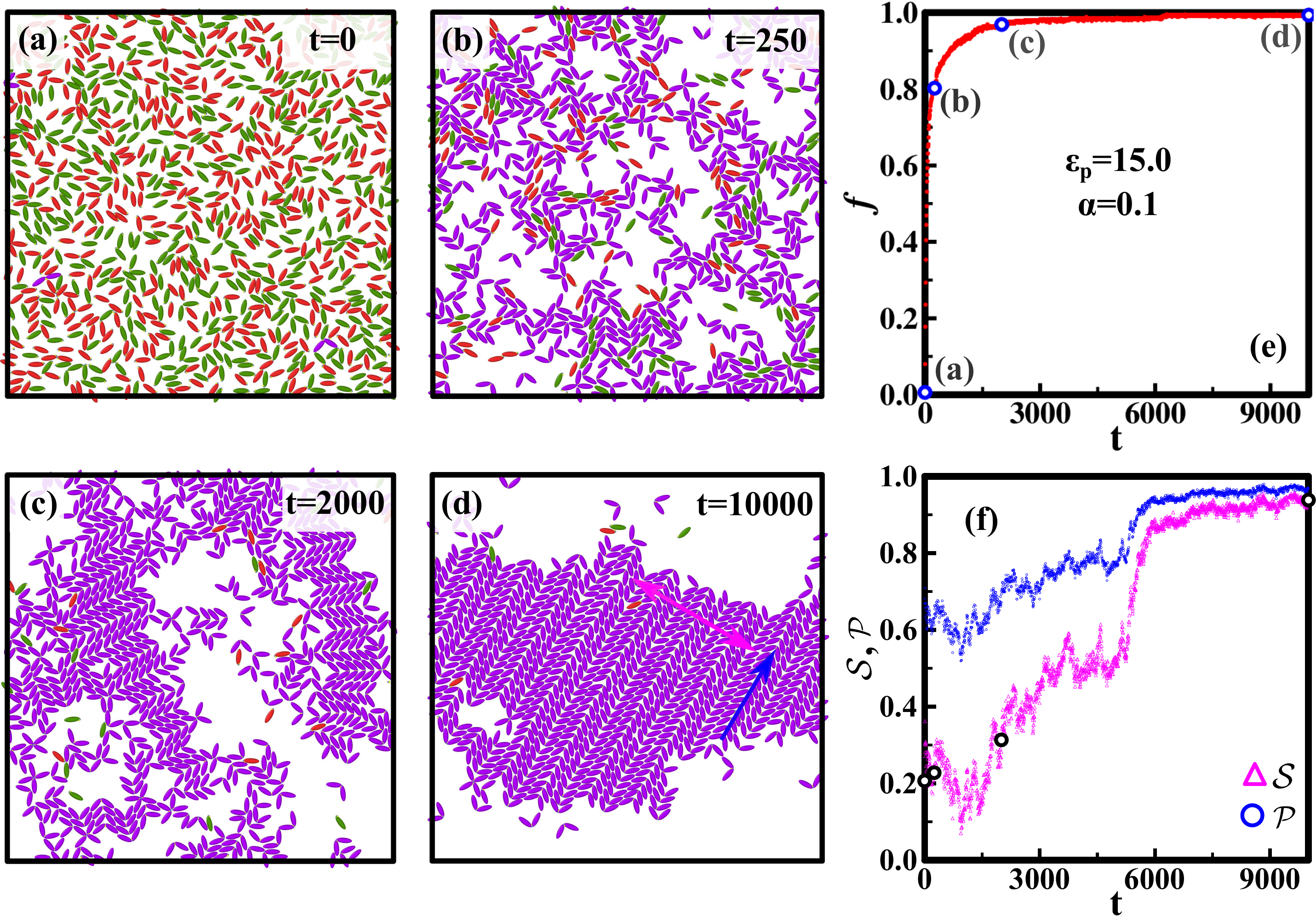}
\caption{Evolution morphologies for $\epsilon_p=15$, $\alpha=0.1$, at $T=1.2$ and $\rho=0.4$ from a typical run at (a) $t=0$, (b) $t=250$, (c) $t=2000$, and (d) $t=10000$. (e) Corresponding BC yields $f$ vs. $t$. The times corresponding to snapshots (a)-(d) are also indicated (open circles). (f) Evolution of the nematic ($\mathcal{S}$) and polar ($\mathcal{P}$) order parameters with simulation time $t$.}
\label{fig:nvt}
\end{figure}

We now turn our attention to the ordering assemblies from the initial monomeric state. Figs.~\ref{fig:nvt}(a)-(d) show the time evolution of the morphology. The corresponding BC yields $f$ are shown in Fig.~\ref{fig:nvt}(e).  
At time $t=0$, when the system is quenched from $T=2$ to $1.2$, the system is in a homogeneous isotropic state where the monomers have not formed any BCs [Fig.~\ref{fig:nvt}(a)]. In the absence of any patches, this would remain an isotropic fluid, given our choice of the GB parameters. Due to the presence of attractive patches, opposite-handed monomers start adhering. The BC yield $f$ rapidly approaches 1 as time increases. An unusual observation from Figs.~\ref{fig:nvt}(c) and (d) is that the evolving ordered state has anti-ferroelectric smectic order. 
We also evaluate the nematic and polar order parameters ($\mathcal{S}$ and $\mathcal{P}$) for the evolving morphology in Fig.~\ref{fig:nvt}(f). Note that the polar direction $\hat{p}$ of the neighbouring BCs in the ordered phase are antiparallel (or staggered). Consequently, we evaluate the polarization by considering only the magnitude of $\hat{p}$ along the direction of polarization ${\bf \vec{p}}$. In other words, we take the absolute value of $(\hat{p}_i\cdot\mathbf{p})$ before averaging over all BCs. As seen in Fig.~\ref{fig:nvt}(f), $\mathcal{S}$ is close to zero at early times as the system is in an isotropic state. As time evolves, there is the emergence of BCs and subsequent liquid crystalline order due to the GB interactions between the BC units. The BCs start forming small locally aligned regions, which then coalesce into larger domains with liquid crystalline order [Fig.~\ref{fig:nvt}(c) and (d)]. Consequently, there is an increase in the nematic and polar order parameters, as seen in Fig.~\ref{fig:nvt}(f). These observations, along with Figs.~\ref{fig:nvt}(e), emphasize the two-stage self-assembly in our model: the BC formation followed by long-range liquid crystalline order in the smectic anti-ferroelectric phase.

\begin{figure}[H]
\centering
\includegraphics[width=1\textwidth]{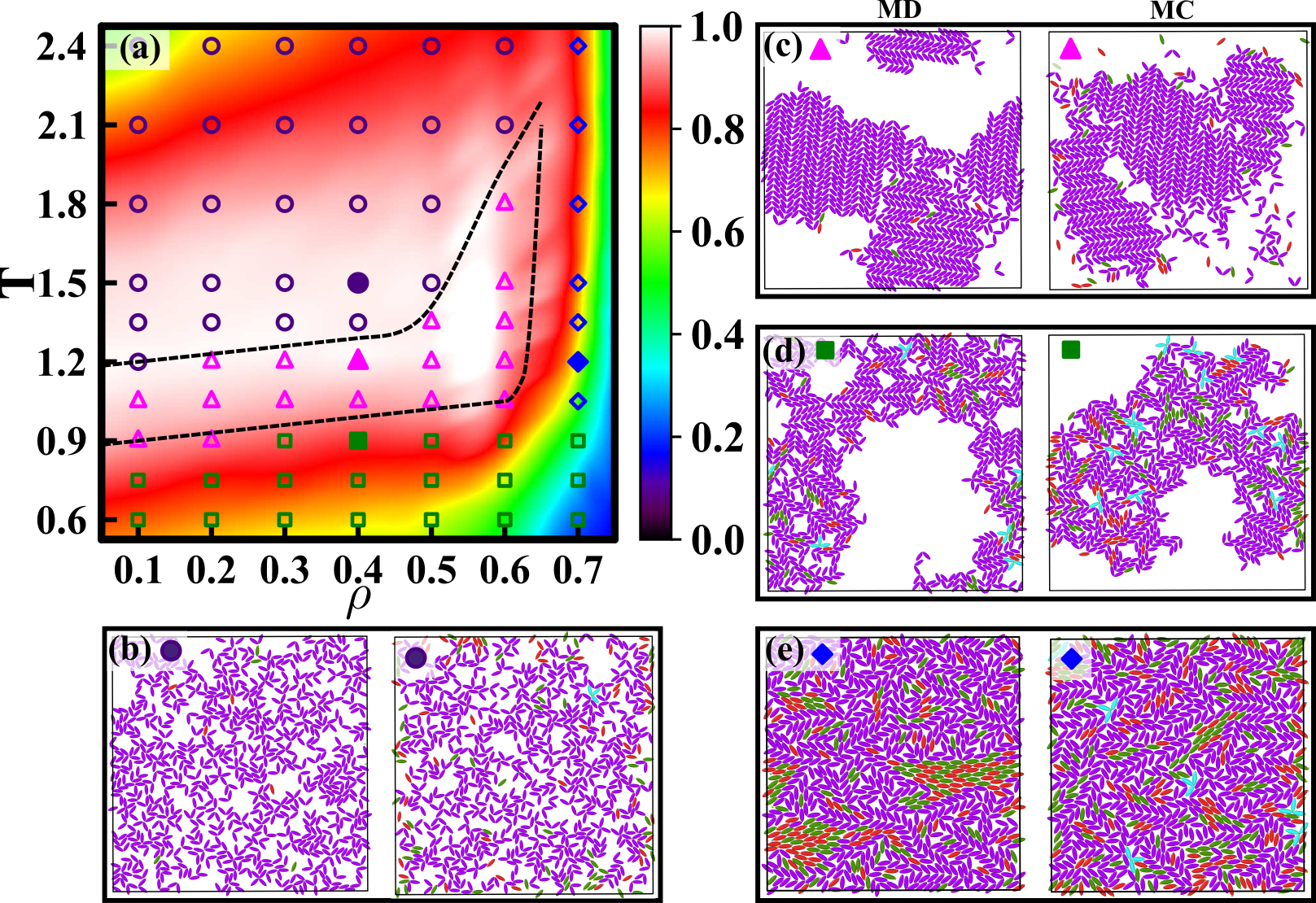}
\caption{(a) $\rho - T$ state diagram for the choice of parameters $\epsilon_p=15$, $\alpha=0.1$, in the NVT ensemble. The symbols correspond to different phases: isotropic (I, indigo circle), anti-ferroelectric smectic (magenta triangle), arrested state due to low-temperature (AS1, green square), and arrested state due to high density (AS2, blue diamond). The background color represents the BC yield $f$ at the given state point. The region between dashed lines corresponds to smectic anti-ferroelectric. Snapshots of typical morphologies obtained from MD (left) and MC (right) simulations at state points marked with filled symbols in panel (a): (b) $T=1.5$, $\rho=0.4$ (I), (c) $T=1.2$, $\rho=0.4$ (smectic anti-ferroelectric), (d) $T=0.9$, $\rho=0.4$ (AS1), and (e) $T=1.2$, $\rho=0.7$ (AS2).}
\label{fig:phase}
\end{figure}

\begin{figure}[H]
\centering
\includegraphics[width=1\textwidth]{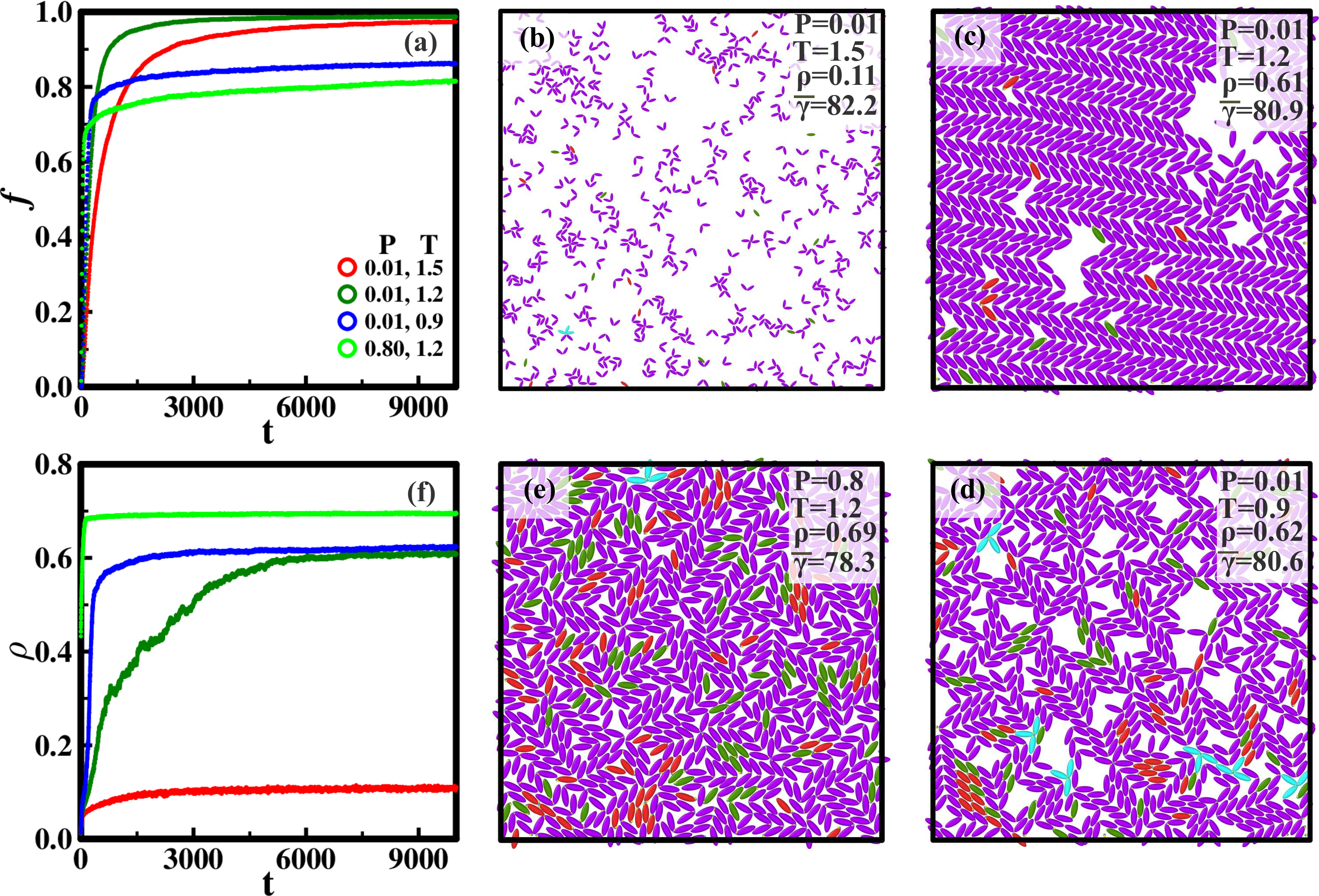}
\caption{The plot of BC yield $f$ vs. $t$ for fixed values of $\epsilon_p=15$ and $\alpha=0.1$ in NPT ensemble for different pressures $P$ and temperatures $T$. Corresponding typical final morphologies at $t=10^4$: (b) $P=0.01$, $T=1.5$; (c) $P=0.01$, $T=1.2$; (d) $P=0.01$, $T=0.9$; (e) $P=0.8$, $T=1.2$. (f) Corresponding density $\rho$ variation with simulation time $t$. The morphologies resemble those in Fig.~\ref{fig:phase} for corresponding values of $\rho$ and $T$.}
\label{fig:pressure}
\end{figure}

The full set of BC states (phases) for a specific choice of model parameters ($\epsilon_p = 15$, $\alpha = 0.1$) in the optimal window is shown in the $\rho - T$ state diagram of Fig.~\ref{fig:phase}(a). The different symbols indicate the different states observed in our simulations: isotropic state (indigo circles), anti-ferroelectric smectic (magenta triangles), arrested or frozen state (green squares) and disordered state (blue diamonds). The background color represents the BC yield. We see that there is a sizeable region in the $\rho-T$ space where the BC yield is greater than $0.9$.
Panels (b) to (e) show typical morphologies obtained from MD (left) and MC (right) simulations for the four states obtained at indicated parameter values of the corresponding filled symbols in the state diagram. In (b),  there is formation of BCs, but the system is in an isotropic state as the Gay-Berne attraction between them is not sufficient to overcome the thermal fluctuations. 
The region between the dashed lines is where we obtain the smectic anti-ferroelectric phase shown in the snapshot (c). For these parameter values, the BCs are fluid enough to self-assemble into large ordered domains. At low temperatures ($T < 0.9$), there are locally aligned regions of BCs, but there is a lack of long-range order as these clusters are frozen. Such a kinetically arrested or frozen state due to low temperatures is shown in panel (d). At very high densities $\rho > 0.6$, the particles lack the free volume to move and reorient, leading to a dense disordered BC state, as shown in panel (e). Similar morphologies are also observed in the NPT ensemble with pressure-tuned to achieve corresponding densities, see Fig.~\ref{fig:pressure}.

\begin{figure}[H]
\centering
\includegraphics[width=1\textwidth]{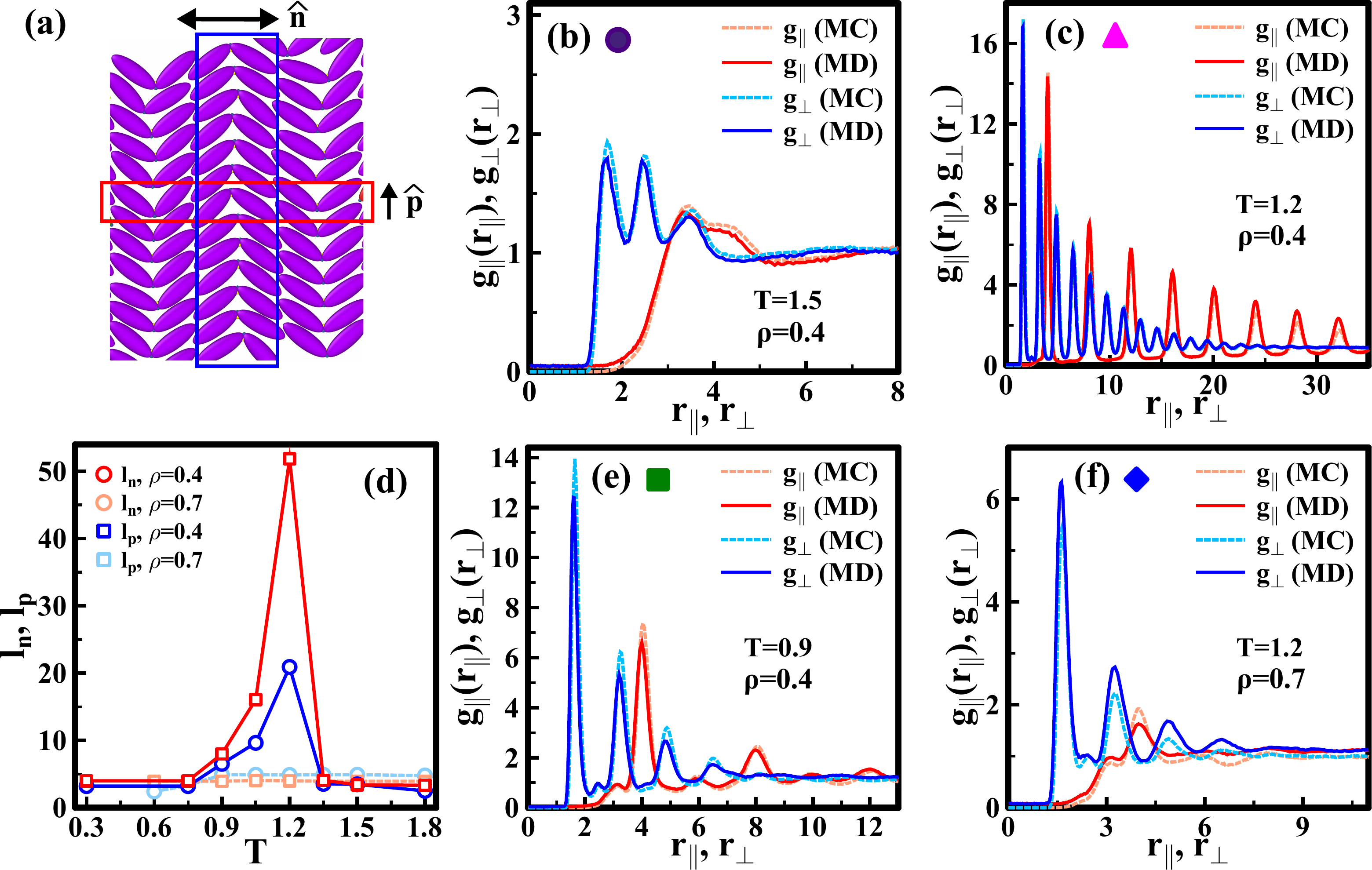}
\caption{(a) Schematic illustration of the evaluation of the directional RDFs along the polar and nematic alignment of the central BC, with correlation lengths $l_n$ along nematic and $l_p$ along polar directions.  Plots of the directional RDFs $g_{\parallel}(r_{\parallel})$ (in red), $g_{\perp}(r_{\perp})$ (in blue) vs. $r_{\parallel}, r_{\perp}$ in the NVT ensemble for various values of $T$ and $\rho$: (b) $T=1.5$, $\rho=0.4$, (c) $T=1.2$, $\rho=0.4$, (e) $T=0.9$, $\rho=0.4$, and (f) $T=1.2$, $\rho=0.7$. All data correspond to the choice of model parameters: $\epsilon_p=15$, $\alpha=0.1$, $\theta_1=5$, and $\theta_2=10$. (d) Correlation lengths $l_n$ and $l_p$ as a function of temperature $T$ at densities $\rho$ = 0.4 and $\rho$ = 0.7.}
\label{fig:pcf}
\end{figure}

In order to characterize the structure of the phases, we evaluate the directional radial distribution functions (RDF) resolved along axes in the local frame of reference of the BCs as described in section~\ref{PCF}. This is schematically illustrated in Fig.~\ref{fig:pcf}(a) where for each BC $i$, we compute the RDF with other BCs along parallel or perpendicular directions to its local alignment $\hat{n}_i$. Only those BCs that are within a finite region perpendicular to the direction of the correlation evaluation are included. The directional RDFs at $\rho=0.4$ and $T=1.5$, corresponding to the isotropic phase, are shown in Fig.~\ref{fig:pcf}(b). We observe only a few peaks in $g_{\perp}(r_{\perp})$ and hardly any in $g_{\parallel}(r_{\parallel})$, indicating that at most, only two or three BCs stack along their short axis. In the case of the smectic anti-ferroelectric phase ($\rho=0.4$ and $T=1.2$), we observe multiple equidistant peaks along either direction, indicating the arrangement of BCs into multiple layers, as shown in Fig.~\ref{fig:pcf}(c). Interestingly, the correlations decay much more slowly in the direction of the long axis of the BCs ($g_{\perp}(r_{\perp})$). In the frozen state at $\rho=0.4$ and $T=0.9$ [Fig.~\ref{fig:pcf}(e)], we find equidistant peaks that decay rapidly, corresponding to short-range correlations along either direction. This is due to the presence of multiple small domains of locally aligned BCs. In the dense disordered state ($\rho=0.7$ and $T=1.2$), we only find short-range correlations along the short axis of the BCs, as seen from Fig.~\ref{fig:pcf}(f). In all cases, our MD and MC results are in agreement with each other. The variation in the correlation lengths along the two directions ($l_n$ and $l_p$) with temperature $T$ is shown in Fig.~\ref{fig:pcf}(d), for two densities ($\rho = 0.4$ and $0.7$). At the higher density, the system is in a dense disordered state, and hence the correlation lengths do not vary with temperature. At the lower density, we see a gradual increase in the correlation lengths with a peak centered around $T = 1.2$, corresponding to the increased smectic anti-ferroelectric alignment, beyond which the correlations decay rapidly as the system enters the high-temperature isotropic phase. The plots also show that the size of the LC domains along the long axis of the BCs is larger than that in the orthogonal direction. The current investigation of parameters belonging to patch-patch interactions provides only isotropic and smectic anti-ferroelectric phases. This is because we have not fully explored the effect of patch positions on the formation of mesoscopic phases. Our preliminary explorations of the $\alpha$-$\theta_1$-$\theta_2$ space reveal a variety of phases, e.g., nematic and splay-bends, through appropriate choice of the model parameters.



\section{Summary and Conclusion}
\label{sec:conclusions}

We have developed a patchy ellipsoid model capable of generating mesoscopic liquid crystalline phases in a hierarchical two-stage self-assembly process, via the formation of BC dimers in the first stage. This is achieved through a binary mixture of ellipses having a pair of complementary and obliquely placed patches with specific combination rules to guide the formation of BCs. The patches are modeled as points on the surface of the ellipses interacting via an inverted Gaussian potential. The second stage of the assembly into ordered liquid phases is governed by the Gay-Berne interactions acting between the ellipses themselves. Our model has three crucial parameters: (i) the patch interaction strength $\epsilon_p$, (ii) the patch interaction range $\alpha$ and (iii) the patch positions $\theta_1$ and $\theta_2$, that control the formation of the BCs, their mean bend angle and flexibility. 

We perform extensive MD and MC simulations to elucidate the model and the effect of the model parameters. Our main observations are summarized below:\\
(a) For fixed patch positions $\theta_1$ and $\theta_2$ and patch range $\alpha$, we observe a window of $\epsilon_p$ values for which the BC yield $f$ is close to $1$. For small values of $\epsilon_p$, the particles do not make stable bonds, whereas for large values of $\epsilon_p$, assemblies of more than two complementary monomers (asters) are formed due to increased stability of single bonds.\\
(b) Similarly, with all other parameters fixed, we obtain high BC yields for a wide range of values of $\alpha$. Although smaller values of $\alpha$ also can produce BCs, we observe low yields in our simulations due to slow kinetics. Larger values of $\alpha$ promote the formation of asters as the monomers can make erroneous bonds due to increased range.\\
(c) For a choice of parameters in the optimal window, we study the evolution of the BC yield and the nematic and polar orders, which show that the assembly into ordered phases occurs in two-stages with the BC formation preceding the long-range ordering.\\
(d) We obtain the state diagram of the model in the $\rho-T$ plane, which exhibits isotropic, smectic anti-ferroelectric and disordered arrested states (due to low temperatures or/and high densities) for the given choice of model parameters. We compute correlation functions and order parameters of these states and show that there is consistency between MD and MC simulations. Further, there is consistency in the morphologies between NVT and NPT ensembles as well, at appropriately chosen parameter values.\\
(e) The patch positions $\theta_1$ and $\theta_2$ can also be varied to generate BCs with different configurations leading to a myriad of interesting liquid crystalline phases, which we shall explore in subsequent works.

The assembly scenarios of these systems are very rich and, so far, the available models for BC entities from patches have not been able to predict {\it all} the observed BCLC phases. Our work sets the stage for these future investigations. In contrast to the molecular BCs often studied in experiments, our model provides a colloidal route to create BCLCs. Generating them at the colloidal scale in experiments is challenging but could have huge potential applications. We believe that our study offers one possible solution for the experimental realization of these interesting fluids.

\section*{Acknowledgements}
AKS gratefully acknowledges the HPC computing facility of IIT Delhi. JT, EB and GK gratefully acknowledge the generous share of CPU time, offered by the Vienna Scientific Cluster (VSC) under project number fs71898. JT acknowledges financial support from the EU-funded doctorate college ENROL. EB acknowledges support from the Austrian Science Fund (FWF) under Proj. No. Y-1163-N27. VB acknowledges SERB (India) for research grants CRG/2022/004871 and CRG/2019/001836. AB acknowledges partial financial support from CRG/2019/001836 during a visit to    IIT Delhi that initiated this work.


\end{document}